\documentclass[leqno,12pt]{article}
\usepackage{amsfonts,latexsym}
\usepackage{amscd}
\usepackage[dvips]{graphicx}

\newcommand{\nequation}{\setcounter{equation}{0}}
\newcommand{\R}{{\Bbb R}}

\newcommand{\ein}{\hspace*{0.65cm}}
\newcommand{\eein}{\hspace*{0.85cm}}

\input epsf
\title{\sc Edge waves along a sloping beach}
\author{Adrian Constantin}
\date{{\small Department of Mathematics, Lund University, P.O. Box
    118, S-22100 Lund, Sweden\\
E-mail: adrian.constantin@math.lu.se}}

\begin{document}
\maketitle

\noindent
We construct a family of explicit rotational solutions to the 
nonlinear governing equations 
for water waves, describing edge waves propagating over 
a plane-sloping beach. A detailed analysis of the edge wave
dynamics and of the run-up pattern is made possible by the use of the
Lagrangian approach to the water motion. A graphical representation of
the edge wave is also presented.\bigskip

---------------------------------------------------------------------------------------------------------

\section{Introduction}

\ein Standing on a gently sloping straight beach, it is a matter of
observation that various waveforms propagate on the surface of the
sea. Among these we find edge waves - water waves that 
progress along the shoreline. 
These waves, often difficult to
visualize (this has, probably, prevented the regarding of this
waveform as important for a long time), 
are coastal trapped, i.e. their amplitude is 
maximal at the shoreline and decays rapidly offshore. They produce 
on the beach beautiful run-up 
patterns (highest points reached by a wave on the beach). Although propagation 
is along the straight shoreline and the waveform is sinusoidal 
in the longshore, these waves 
are not one-dimensional (Lighthill 1978). 

While they were 
originally considered 
to be a curiosity (Lamb 1932), edge waves are now recognized to 
play a significant role in nearshore hydrodynamics. For shallow 
beaches empirical evidence shows that incident storm
waves loose most of their energy trough wave breaking by the time
they reach the shore. After breaking offshore, as the waves progress to
shallower water, their height decreases reaching its least value at
the shoreline. Since storms often result in pronounced shoreline
erosion, the surf zone water processes with the onset of a storm are
dominated by wave conditions other than the incident waves - role
attributed by oceanographers to the edge waves ; see Komar
(1998). There are other instances when edge waves are of
significance. For example, processed data from the water waves
created by an earthquake occuring in April 1992 in the ocean floor
near the Californian coast show that two distinct wave packets (both 
directly generated in the nearshore by the vertical motion
of the ocean bottom) were
recorded at a coastal station about $150\,km$ far from the epicentre. At
first, less than an hour after the occurence of the earthquake, 
relatively fast-moving swell with amplitude around $15\,cm$ striked 
from offshore. About two hours after the swell has
subsided, relatively slow-moving edge
waves with amplitudes around $50\,cm$ (Boss et al. 1995) arrived. 
Measurements performed on this occasion confirm the rapid decay of
the amplitude of the edge waves: at an offshore distance of $12\,km$
the amplitude is reduced to $10\,\%$ of its maximal value 
(attained at the shoreline). Let us also mention that it has been
observed (see Evans 1988) that hurricanes travelling approximately
parallel to a nearby coastline sometimes give rise to edge
waves. Interestingly, edge waves can in fact be generated directly in
a laboratory wave tank (Yeh 1985).

The edge wave phenomenon has been
extensively studied and discussed in the mathematical 
literature within the framework of linear 
theory. Due to the small
displacements associated with these waves, the governing
equations for water waves or the shallow-water equations 
are linearized (Minzoni \& Whitham 1977 showed that both
approximations are equally consistent) and this simplification 
permits a thorough analysis. We refer to Ehrenmark (1998) 
for an excellent up-to-date survey. 

Despite the fact that the 
linearizing approximation lacks rigorous mathematical 
justification, it has been used with considerable success as a large variety
of theoretical studies are confirmed in experimental contexts. 
The investigation of nonlinear edge
waves can be seen as a natural extension to the linear theory. 
Whitham (1976)
showed the existence of irrotational 
weakly nonlinear edge waves that propagate
parallel to the shore using a formal Fourier series expansion for the full
water-wave theory. A study of properties of nonlinear progressive edge
waves based on the fact that the evolution is described by the
nonlinear Schr\"odinger equation was performed by Yeh (1985). This
paper describes an alternative approach; the main impetus for the
results reported here comes from the belief that 
the need for a more rigorous theory remains thoroughly
justified. A quest for an explicit edge wave solution for the governing
equations appears to be of interest since the structure of the edge
waves obscures their visual observation; moreover, a solution in
closed mathematical form  provides a background
against which certain features which have been observed (and
predicted) can be checked. It turns out
that the deep water wave solution discovered by Gerstner (1809), 
can be adapted 
to construct edge waves propagating along a plane-sloping beach. This
possibility was pointed out by Yih (1966) but the treatement therein,
in essence followed also by Mollo-Christensen (1982), 
provides only an implicit form for the
free water surface. We present a procedure by which 
exact edge wave solutions to the full 
water-wave equations are obtained. The closed form of
the solution in Lagragian (material) coordinates permits us to provide clear
illustrations of the structure of these edge waves. From an examination
of the solution we also obtain the run-up pattern, an attractive
feature being the occurence of cusps - these shapes are confirmed in
both field and laboratory evidence (see Komar 1998). Thus, we 
establish with rigour the existence of rotational nonlinear edge
waves, unraveling the
detail structure of the wave pattern.

\section{The edge wave}
\nequation

We take a plane beach and adopt a coordinate system as shown below,
with the shoreline the $x$-axis, and the still sea in the region
$$R=\{(x,y,z):\ x \in \R,\, y \le b_0,\, 0 \le z \le
(b_0-y)\,\tan\alpha\}$$
for some $b_0 \le 0$; here $\alpha \in (0,\displaystyle\frac{\pi}{2})$
defines the uniform slope.

\begin{center}
\includegraphics[width=10cm]{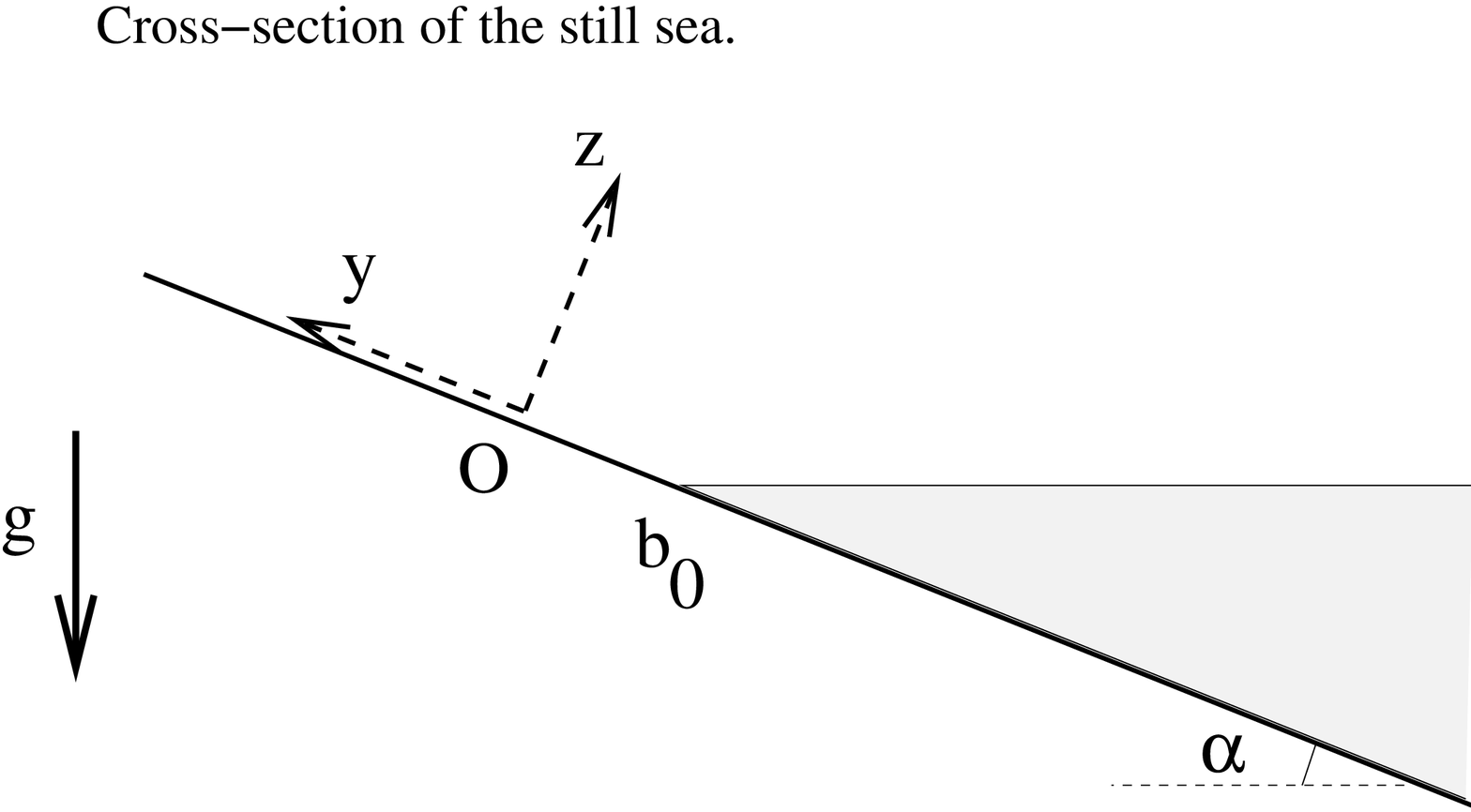}
\end{center}

\noindent
Let $u=(u_1,u_2,u_3)$ be the velocity field and let us recall
(e.g. Crapper 1984) the governing equations for the propagation of 
gravity water waves
when ignoring viscous effects. Homogeneity (constant density 
$\rho$) is a good 
approximation for water (see the numerical data in Lighthill 1978) 
so that we have the equation of mass conservation in the 
form
\begin{equation}
\frac{\partial u_1}{\partial x}+\frac{\partial u_2}{\partial y}+
\frac{\partial u_3}{\partial z}=0.
\end{equation} 
The equation of motion is Euler's equation,
\begin{equation}
  \left\{
  \begin{array}{l}
\displaystyle\frac{D u_1}{D t}=-\frac{1}{\rho}\,\frac{\partial P}{\partial x},
\\[0.5cm]
\displaystyle\frac{D u_2}{D t}=-\frac{1}{\rho}\,
\frac{\partial P}{\partial y}-g\,\sin\alpha,
\\[0.5cm]
\displaystyle\frac{D u_3}{D t}=-\frac{1}{\rho}\,
\frac{\partial P}{\partial z}-g\,\cos \alpha.
\end{array} \right.  
\end{equation}
where $P(t,x,y,z)$ denotes the pressure, $g$ is the gravitational
acceleration constant and $D/Dt$ is the material time
derivative, $\displaystyle\frac{D f}{D t}=\frac{\partial f}{\partial
  t}+u_1\,\frac{\partial f}{\partial x}+u_2\,\frac{\partial f}{\partial
  y}+u_3\,\frac{\partial f}{\partial
  z}$, expressing the rate of change of the quantity $f$ associated
with the same fluid particle as it moves about. The boundary
conditions which select the water-wave problem from all other
possible solutions of the equations (1)-(2) are (see Johnson 1997):

(i) the dynamic boundary condition $P=P_0$ at the free surface, where
$P_0$ is the constant atmospheric pressure, decouples the motion of 
the air from that of the water;

(ii) the kinematic boundary condition at the free surface expresses 
the fact that the same particles always form the free water surface;

(iii) the kinematic boundary condition at the bottom, requiring the
normal velocity component at the bed to be zero so that it is impossible
for water to penetrate.

The general description of the propagation of a water wave 
is encompassed by the equations (1)-(2) and the three boundary
conditions (i)-(iii), a distinctive feature being 
that the free surface is not known and must be 
determined as part of the solution.

We adopt the Lagrangian point of view by following the evolution of
individual water particles. We suppose that the position of a particle
at time $t$ is given by 
\begin{equation}
  \left\{
  \begin{array}{l}
x=a-\displaystyle\frac{1}{k}\,e^{k(b-c)}\sin\Bigl(ka+\sqrt{gk\,\sin\alpha}\
t\Bigr),
\\[0.5cm]
y=b-c+\displaystyle\frac{1}{k}\,e^{k(b-c)}\cos\Bigl(ka+\sqrt{gk\,\sin\alpha}\
t\Bigr),
\\[0.5cm]
z=c+c\,\tan\alpha+\displaystyle\frac{\tan\alpha}{2k}\,e^{2kb_0}
\Bigl(1-e^{-2k\,c\,(1+\cot \alpha)}\Bigr),
\end{array} \right.  
\end{equation}
where $k>0$ is fixed. It should be pointed out that the quantities
$a,b,c$ do not stand for the initial coordinates of a particle, they
are simply labeling variables serving to identify a particle. We may
think of them as parameters which fix the position of a particular
particle before the passage of the wave (in still water), despite the
fact that the wave is not developing from the still state - otherwise,
the flow would be irrotational in view of Helmholtz's theorem (see
Johnson 1997) but its vorticity is
nonzero (see the last section). Let us explain the origin of
(3). Gerstner (1809; see Constantin 2001) gave the only known
nontrivial explicit
solution to the full water-wave equations, showing that the
two-dimensional particle motion ($a \in \R,\, b \le b_0 \le 0,\, k>0$)
$$t \mapsto \Bigl( a+\frac{1}{k}\,e^{kb}\sin(ka+\sqrt{gk}\ t), 
\ b-\frac{1}{k}\,e^{kb}\cos(ka+\sqrt{gk}\ t)\Bigr)$$
represents waves of finite amplitude in water of infinite depth. This
suggests that it might be possible to construct an edge wave using an 
approach similar to the one for a Gerstner
wave field. While the theoretical correctness of this conclusion 
was established by Yih (1966) and Mollo-Christensen (1982), the
outcome in both treatements was an implicit form the water's free
surface which makes the obtained waveform graphically and
computationally inaccessible. The closed form (3) provides the full
details of the edge wave motion without considerable labour. 

Our aim is to prove that the motion (3) is dynamically possible and
that we can associate to it an expression for the hydrodynamical
pressure $P$ such that the governing equations and boundary conditions
are all satisfied. The resulting free surface of the water will be the
edge wave we are looking for.

The map (3) is a
diffeomorphism from the still water region $R$ to the water region
bounded below by the rigid bed $\{z=0\}$ and above by the free water
surface parametrized by 
\begin{equation}
  \left\{
  \begin{array}{l}
x=a-\displaystyle\frac{1}{k}\,\displaystyle{e^{kb(1+\tan\alpha)-kb_0\tan\alpha}
\sin\Bigl(ka+\sqrt{gk\,\sin\alpha}\
t\Bigr)},
\\[0.5cm]
y=b(1+\tan\alpha)-b_0\tan\alpha+\displaystyle\frac{1}{k}\,e^{kb(1+\tan\alpha)-
kb_0\tan\alpha}\cos\Bigl(ka+\sqrt{gk\,\sin\alpha}\
t\Bigr),
\\[0.5cm]
z=(b_0-b)\,(1+\tan\alpha)\tan\alpha+\displaystyle\frac{\tan
\alpha}{2k}\,e^{2kb_0}
\Bigl(1-e^{2k(b-b_0)\,(1+\tan \alpha)}\Bigr),
\end{array} \right.  
\end{equation}
with $a \in \R$, $b \le b_0$, and $t \ge 0$. Indeed, observing that
$(a,b,c) \mapsto (a,b-c,c)$ defines a diffeomorphism of $\R^3$, it is
enough to show that the map 
$$\left(
\begin{array}{c}
a \\
b'\\
c\\
\end{array} \right) \mapsto \left(
\begin{array}{c}
a-\displaystyle\frac{1}{k}\,\displaystyle{e^{kb'}
\sin\Bigl(ka+\sqrt{gk\,\sin\alpha}\
t\Bigr)}
\\[0.5cm]
b'+\displaystyle\frac{1}{k}\,e^{kb'}\cos\Bigl(ka+\sqrt{gk\,\sin\alpha}\
t\Bigr)
\\[0.5cm]
c+c\,\tan\alpha+\displaystyle\frac{\tan\alpha}{2k}\,e^{2kb_0}
\Bigl(1-e^{-2k\,c\,(1+\cot \alpha)}\Bigr)
\end{array} \right)$$
is a diffeomorphism on $\R \times \R_- \times \R_+$. To see this,
observe that the third coordinate depends only on $c$, being an
increasing function $f(c)$ with $f(0)=0$ and $\lim_{c \to
  \infty}f(c)=\infty$. Therefore, we may slice $\R \times \R_- \times
\R_+$ by planes parallel to the plane $c=0$ and in each such plane the
particle motion is precisely that of a Gerstner wave field (see 
Constantin 2001). This proves that (3) is a
diffeomorphism and it is easy to identify the boundary of the image of
the region $R$ under it.

The Lagrangian form of the equation of continuity (that is, the 
volume-preserving property of the flow) is fulfilled since the value
of the Jacobian of the map (3) is independent of time. This, together
with the previously proved fact that (3) defines at any fixed time a
diffeomorphism, shows that the motion described by (3) is dynamically
possible. To complete the proof that (3) describes the water motion induced by
a gravity wave, we have to check Euler's equation (2) and the boundary
conditions (i)-(iii) for a suitably defined value of the
hydrodynamical pressure.

The acceleration of a particular water particle is
$$\frac{Du}{Dt}=\Bigl(
g\,\sin\alpha\,e^{k(b-c)}\sin(ka+\sqrt{gk\sin\alpha}\ t)\,,\, 
-g\,\sin\alpha\,e^{k(b-c)}\cos(ka+\sqrt{gk\sin\alpha}\ t)\,,\,
0\Bigr)$$
so that the equation of motion (2) is
$$\left\{
  \begin{array}{l}
\displaystyle\frac{\partial P}{\partial 
x}=-\rho\,g\sin\alpha\,e^{k(b-c)}\sin(ka+\sqrt{gk\sin\alpha}\ t),
\\[0.5cm]
\displaystyle\frac{\partial P}{\partial 
y}=\rho\,g\sin\alpha\,e^{k(b-c)}\cos(ka+
\sqrt{gk\sin\alpha}\ t)-\rho\,g\sin\alpha, 
\\[0.5cm]
\displaystyle\frac{\partial P}{\partial 
z}=-\rho\,g\,\cos \alpha.
\end{array} \right.$$
Passing to Lagrangian coordinates, we obtain the system
$$\left\{
  \begin{array}{l}
\displaystyle\frac{\partial P}{\partial 
a}=0,
\\[0.5cm]
\displaystyle\frac{\partial P}{\partial 
b}=\rho\,g\sin\alpha\,e^{2k(b-c)}-\rho\,g\sin\alpha, 
\\[0.5cm]
\displaystyle\frac{\partial P}{\partial 
c}=-\rho\,g\sin\alpha\,e^{2k(b-c)}-\rho\,g\,\cos \alpha+
\rho\,g\cos\alpha\,(1+\tan\alpha)\,e^{2kb_0}e^{-2kc(1+\cot\alpha)},
\end{array} \right.$$
with the solution
$$P=P_0+\frac{\rho g \sin\alpha}{2k}\, e^{2k(b-c)}- 
\rho g\Bigl(c\,\cos\alpha+(b-b_0)\,\sin\alpha\Bigr)- 
\frac{\rho g \sin\alpha}{2k}\, 
e^{-2kc(1+\cot \alpha)}e^{2kb_0}.$$
The obtained hydrodynamical pressure has the same value for any given
particle as it moves about. At the free surface $c=(b_0-b)\tan\alpha$
we have $P=P_0$ so that the dynamic boundary condition (i) is
satisfied. The kinematic boundary condition at the free surface, (ii),
holds as at any instant the free surface (4) is the image under (3) of
the still water surface $\{c=(b_0-b)\tan\alpha:\ b \le b_0\}$. That
there is no velocity normal to the sloping shore - this takes care of
the boundary condition (ii) - is obvious, because at $z=0$ we have
$c=0$ and the motion (3) is planar, without any velocity component in
the direction of $z$. The proof of the fact that (3) is an explicit
solution to the governing equations for water waves on a plane-sloping
beach is complete.

\section{Discussion}

\ein We have presented an exact edge wave solution to the full
water-wave problem, the graphical depiction of which is a fairly
easy exercise. Let us now emphasize some of its significant properties. 
We present some simple observations which will provide a
comprehensive description of this nonlinear wave and the
particle motion it induces below the water surface.    

The wavelength in the longshore direction $\lambda=2\pi/k$ is related
to the wave frequency $\omega$ by
$$\omega^2=gk\,\sin\,\alpha,$$
while the wave period is 
$$T=\frac{2\pi}{\sqrt{ gk\,\sin\,\alpha} }.$$
We easily infer that
$$\lambda=\frac{gT^2}{2\pi}\,\sin\alpha,$$
so that the length of the edge wave is strongly dependent on its 
period and to a smaller degree on the beach slope. 
The phase velocity $U$ of the edge wave (4) is given by
\begin{equation}
U=\sqrt{\frac{g\sin\alpha}{k}},
\end{equation}
a fact consistent with the observation that if the bottom is flat
($\alpha=0$), then $U=0$ and no edge wave exists. The dispersion
relation (5) for edge waves is obtained (see Johnson 1997) within the
confines of the formal linear approximation to the governing
equations, but in our case the relation is derived rigorously as a
byproduct of (3). 

From (3) it is clear that any water particle
describes circles as the edge wave passes - all these
circles lie in planes parallel to the sloped bottom. The radius
$\displaystyle\frac{1}{k}\,e^{k(b-c)}$ of the circle described
counterclockwise by a
particle is maximal for
the particles at the shoreline (that is, for $b=b_0,\,c=0$).

As pointed out in the previous section, the motion of the water 
body induced by the passage of the edge wave
(4) is rotational. The vorticity of the water flow defined by (3) is
$$\hbox{curl}\, u=\Bigl(-\,\frac{\partial u_2}{\partial z},\ 
\frac{\partial u_1}{\partial z},\ \frac{\partial u_2}{\partial x}-
\frac{\partial u_1}{\partial y}\Bigr),$$
by the vanishing of $u_3$. Computing the inverse of 
the Jacobian matrix of the diffeomorphism
(3) as
$$\left(
\begin{array}{ccc}
\frac{1+\exp[k(b-c)]\,\cos\,k(a+Ut)}{1-\exp[2k(b-c)]} & 
\frac{\exp[k(b-c)]\,\sin\,k(a+Ut)}{1-\exp[2k(b-c)]} & 0\\[0.5cm]
\frac{\exp[k(b-c)]\,\sin\,k(a+Ut)}{1-\exp[2k(b-c)]} & 
\frac{1-\exp[k(b-c)]\,\cos\,k(a+Ut)}{1-\exp[2k(b-c)]} & 0\\[0.5cm]
0 & \frac{1}{(1+\tan\alpha)\,(1+\exp[2k(b_0-c-c\cot\alpha)])} & 
\frac{1}{(1+\tan\alpha)\,(1+\exp[2k(b_0-c-c\cot\alpha)])} \\
\end{array} \right),$$
a straightforward calculation yields the expression of the vorticity 
$$ \hbox{curl}\,u=-\,\Bigl(0,\,0,\, 
\frac{2kU}{1-e^{2k(b-c)}}\ e^{2k(b-c)}\Bigr)$$
at a particle whose parameters are $(a,b,c)$. Therefore the vorticity
is in the opposite sense to the revolution of the particles in their
circular orbits, decreasing rapidly with distance from the
shoreline/bed. Observe that, despite the fact that the flow (3)
is not two-dimensional, the vorticity of each individual water
particle is conserved as the particle moves about.

The run-up pattern is obtained by setting $z=0$ in (4); this forces
$b=b_0$ so that we have 
\begin{equation}
  \left\{
  \begin{array}{l}
x=a-\displaystyle\frac{1}{k}\,\displaystyle{e^{kb_0}
\sin\Bigl(ka+\sqrt{gk\,\sin\alpha}\
t\Bigr)},
\\[0.5cm]
y=b_0+\displaystyle\frac{1}{k}\,e^{kb_0}\cos\Bigl(ka+\sqrt{gk\,\sin\alpha}\
t\Bigr),
\\[0.5cm]
z=0,
\end{array} \right.  
\end{equation}
with $a \in \R$. The above formula represents the parametrization of a
smooth trochoid (if $b_0<0$) or of a cycloid with upward cusps (if
$b_0=0$); it also
  explains why we imposed the condition $b_0 \le 0$ as otherwise we
  would obtain a self-intersecting curve.

\begin{center}
\includegraphics[width=10cm]{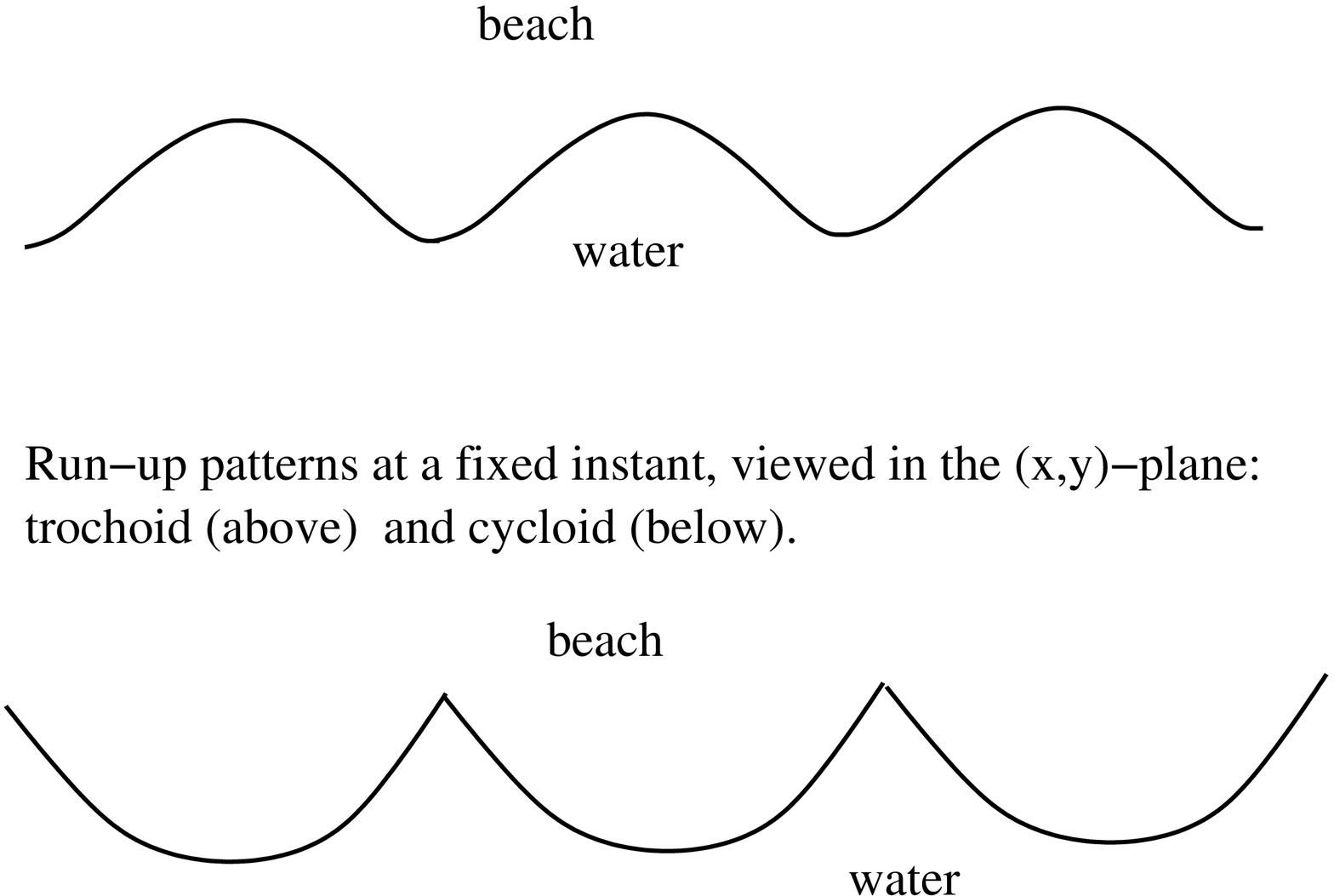}
\end{center} 

\noindent
The cusp formation by some edge waves was clearly demonstrated in wave-tank
experiments and found by field measurements on the ocean beach (see
Komar 1998).

Another aspect of interest is the amplitude of the edge
wave. To determine the
elevation with respect to the reference plane 
$$\{z=\frac{\sin
\alpha}{2k}\,e^{2kb_0}+(b_0-y)\,\tan\alpha:\
y \le b_0\},$$
we compute the distance of a point $(x,y,z)$ lying on the free surface
(4) to this plane, 
$$d=z\,\cos\alpha+(y-b_0)\,\sin\alpha-\frac{\sin\alpha}{2k}
\,e^{2kb_0},$$
with the understanding that positive/negative values on the right-hand
side mean that the point lies above/below the plane. Since (with $b
\le b_0$)
\begin{equation}
d=\frac{\sin\alpha}{2k}\,\Bigl(2\,e^{kb(1+\tan\alpha)-kb_0\tan\alpha}
\cos(ka+\sqrt{gk\,\sin\alpha}\
t)-\,e^{2kb(1+\tan\alpha)-2kb_0\tan\alpha}\Bigr),
\end{equation}
we see that the amplitude of the edge wave decays exponentially away
from the shoreline (as $b \to -\infty$). The same conclusion
is reached by a formal linear approximation (see Whitham 1979) and explains
why edge waves are called ``trapped waves''.

As expected - and ensured by (7), the amplitude of 
the edge wave varies with the parameters
$a \in \R,\ b \le b_0$. From (7) we also infer that, at a fixed $b \le b_0$, 
the crests and troughs correspond to the
maximal/minimal values of $\cos(ka+\sqrt{gk\,\sin\alpha}\
t)$. At a fixed time $t \ge 0$, we
obtain the crest curves (with $m \in \Bbb Z$ fixed and 
$b \le b_0$ playing the role of a running parameter)
$$\left\{
  \begin{array}{l}
x=\displaystyle\frac{2m\pi}{k}-\displaystyle\frac{1}{k}\,
\sqrt{gk\,\sin\alpha}\ t,
\\[0.5cm]
y=b(1+\tan\alpha)-b_0\tan\alpha+\displaystyle\frac{1}{k}\,e^{kb(1+
\tan\alpha)-kb_0\tan\alpha},
\\[0.5cm]
z=(b_0-b)(1+\tan\alpha)+\displaystyle\frac{\tan\alpha}{2k}\,e^{2kb_0}
(1-e^{2k(b-b_0)(1+\tan\alpha)}),
\end{array} \right.$$
and the trough curves 
$$\left\{
  \begin{array}{l}
x=\displaystyle\frac{(2m+1)\pi}{k}-\displaystyle\frac{1}{k}\,
\sqrt{gk\,\sin\alpha}\ t,
\\[0.5cm]
y=b(1+\tan\alpha)-b_0\tan\alpha-\displaystyle\frac{1}{k}\,e^{kb(1+
\tan\alpha)-kb_0\tan\alpha},
\\[0.5cm]
z=(b_0-b)(1+\tan\alpha)\tan\alpha+\displaystyle\frac{\tan\alpha}{2k}\,e^{2kb_0}
(1-e^{2k(b-b_0)(1+\tan\alpha)}).
\end{array} \right.$$
Note that for both the crest and trough curves
the value of $x$ (at a given time) is fixed: standing at that location
and looking towards the sea, these curves, orthogonal to the
shoreline, are fully visible at certain instants. Indeed, 
taking into account the fact
that on a crest/trough curve the deviation from the reference plane is
$$^+_-\frac{\sin\alpha}{2k}\,\Bigl(2\,e^{kb(1+\tan\alpha)-kb_0\tan\alpha}
-\,e^{2kb(1+\tan\alpha)-2kb_0\tan\alpha}\Bigr),\qquad b \le b_0,$$
in view of (7), the monotonicity of the right-hand side shows that the
deviation becomes smaller (in absolute value) 
with the distance from the shore. This feature can be easily
recognized in the graphical representation of the edge wave given
below.

\bigskip

\begin{center}
\includegraphics[width=15cm]{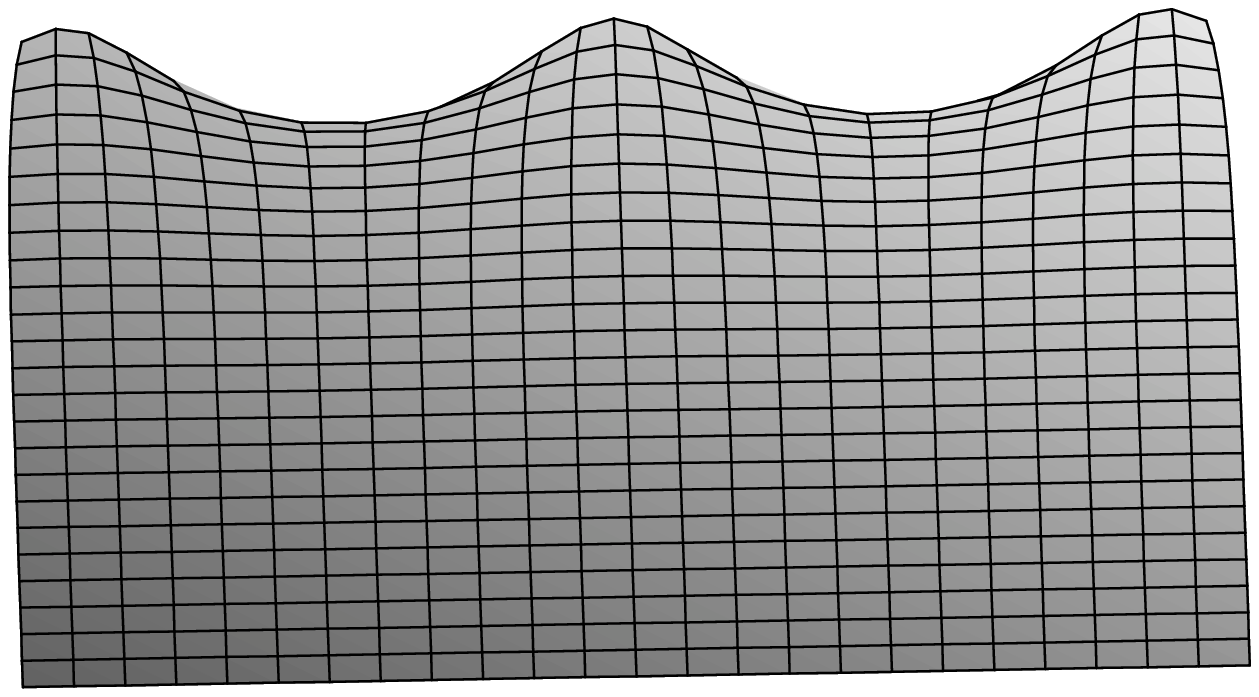}
\end{center} 

\noindent

\eein{\small The edge wave viewed from offshore. The sinusoidal
  longshore structure and the 

\eein exponential offshore decay in amplitude
  are clearly visible.}

\vfill\eject

\centerline{REFERENCES}\bigskip

\noindent
{\sc Boss, E., Gonzalez, F., Sakate, K. \& Mofjeld, H. 1995}  Edge wave
and non-

\eein trapped modes of the 25 April 1992 Cape Mendocino
tsunami. {\it Pure Appl. 

\eein Geophys.} {\bf 144}, 409-426.\smallskip

\noindent
{\sc Constantin, A. 2001} On the deep water wave motion. {\it
  J. Phys. A} {\bf 34}, 1405-1417.\smallskip

\noindent
{\sc Crapper, G. 1984} {\it Introduction to Water Waves}. Ellis
Horwood, Chichester.\smallskip

\noindent
{\sc Ehrenmark, U. 1998} Oblique wave incidence on a plane beach: the
classical problem 

\eein revisited. {\it J. Fluid Mech.} 
{\bf 368}, 291-319.\smallskip

\noindent
{\sc Evans, D. 1988} Mechanisms for the generation of edge waves over a
sloping beach. 

\eein {\it J. Fluid Mech.} {\bf 186}, 379-391.\smallskip

\noindent
{\sc Gerstner, F. 1809} Theorie der Wellen samt einer daraus
abgeleiteten Theorie der 

\eein Deichprofile. 
{\it Ann. Physik} {\bf 2}, 412-445.\smallskip 

\noindent
{\sc Johnson, R. 1997} {\it A Modern Introduction to the Mathematical Theory
  of Water 

\eein Waves}. Cambridge University Press.\smallskip

\noindent
{\sc Komar, P. 1998} {\it Beach Processes and
  Sedimentation}. Prentice-Hall, Inc. \smallskip 

\noindent
{\sc Lamb, H. 1932} {\it Hydrodynamics}. Cambridge University Press.\smallskip

\noindent
{\sc Lighthill, J. 1978} {\it Waves in Fluids}. Cambridge 
University Press.\smallskip

\noindent
{\sc Minzoni, A. \& Whitham, G. B. 1977} On the excitation of edge
waves on beaches. 

\eein {\it J. Fluid Mech.} {\bf 79}, 273-287.\smallskip

\noindent
{\sc Mollo-Christensen, E. 1982} Allowable discontinuities in a Gerstner
wave field. 

\eein {\it Phys. Fluids} {\bf 25}, 586-587.\smallskip

\noindent
{\sc Whitham, G. B. 1976} Nonlinear effects in edge waves. 
{\it J. Fluid Mech.} {\bf 74}, 353-368.\smallskip

\noindent
{\sc Whitham, G. B. 1979} {\it Lectures on Wave Propagation}. Springer
Verlag, Berlin.\smallskip

\noindent
{\sc Yeh, H. 1985} Nonlinear progressive edge waves: their instability
and evolution. {\it J. 

\eein Fluid Mech.} {\bf 152}, 479-499.\smallskip

\noindent
{\sc Yih, C. 1966} Note on edge waves in a stratified fluid. {\it
  J. Fluid Mech.} {\bf 24}, 765-767.\smallskip

\end{document}